\documentclass[twoside,english,reprint, aps, prl, superscriptaddress, amsfonts, amssymb, amsmath, a4paper]{revtex4-1}
\usepackage{lmodern}

\usepackage[T1]{fontenc}
\usepackage[latin9]{inputenc}
\usepackage{color}
\usepackage{babel}
\usepackage{amsmath}
\usepackage{amssymb}
\usepackage{graphicx}
\usepackage[unicode=true,
 bookmarks=false,
 breaklinks=true,pdfborder={0 0 0},pdfborderstyle={},backref=false,colorlinks=true]
 {hyperref}
\hypersetup{pdftitle={Relaxation Dynamics of Spin Qubits in Quantum Dots},
 pdfauthor={Dan Cogan}}

\makeatletter

\providecommand{\tabularnewline}{\\}

\makeatother

\begin{document}
\title{The coherence of quantum dot confined electron- and hole-spin in low
external magnetic field}
\author{Dan Cogan}
\affiliation{The Physics Department and the Solid State Institute, Technion\textendash Israel
Institute of Technology, 3200003 Haifa, Israel}
\author{Zu-En Su}
\affiliation{The Physics Department and the Solid State Institute, Technion\textendash Israel
Institute of Technology, 3200003 Haifa, Israel}
\author{Oded Kenneth}
\affiliation{The Physics Department and the Solid State Institute, Technion\textendash Israel
Institute of Technology, 3200003 Haifa, Israel}
\author{David Gershoni}
\email{dg@physics.technion.ac.il}

\affiliation{The Physics Department and the Solid State Institute, Technion\textendash Israel
Institute of Technology, 3200003 Haifa, Israel}
\begin{abstract}
We investigate experimentally and theoretically the temporal evolution
of the spin of the conduction band electron and that of the valence
band heavy hole, both confined in the same semiconductor quantum dot.
In particular, the coherence of the spin purity in the limit of a
weak externally applied magnetic field, comparable in strength to
the Overhauser field due to fluctuations in the surrounding nuclei
spins. We use an all-optical pulse technique to measure the spin evolution
as a function of time after its initialization. We show for the first
time that the spin purity performs complex temporal oscillations which
we quantitatively simulate using a central spin model. Our model encompasses
the Zeeman and the hyperfine interactions between the spin and the
external and Overhauser fields, respectively. Our novel studies are
essential for the design and optimization of quantum-dot-based entangled
multi-photon sources. Specifically, cluster and graph states, which
set stringent limitations on the magnitude of the externally applied
field.
\end{abstract}
\maketitle
\global\long\def\ket#1{\left|#1\right\rangle }%
\global\long\def\im{\operatorname{Im}}%
\global\long\def\bra#1{\left\langle #1\right|}%

Semiconductor quantum-dot-based devices are currently the most viable
technology for generating quantum light for future quantum information
and network applications. They are easily incorporated into electro-optical
classical components \citep{Reithmaier2004,Yoshie2004,Arcari2014,Senellart2017,Najer2019}
while demonstrating unparalleled efficiency, photon extraction rates
\citep{Somaschi2016,Senellart2017,Najer2019}, and high quality, nearly
transform-limited photon indistinguishability \citep{Kuhlmann2013,Somaschi2016,Ding2016,Najer2019}.
Furthermore, a quantum dot (QD) acts like a single atom emitter and
thus can be a source of entangled photons via its spontaneous emission
\citep{Kim1999,Akopian2006,Winik2017}.

One particularly important resource for quantum communication is a
multi-photonic entangled state - a cluster or a graph state \citep{Zwerger2012,Munro2012,Azuma2015,Buterakos2017}.
Schemes that rely on entangling the spin degree of freedom of QD confined
charge carriers with the polarization degree of freedom of the photons
that QD emit have been proposed \citep{Lindner2009,Economou2010}
and experimentally demonstrated \citep{Schwartz2016}. These schemes
use a single spin precessing in an externally applied magnetic field
while being driven by a sequence of precisely timed short laser pulses.
Upon each excitation of the QD spin, a single photon is spontaneously
emitted, and the photon's polarization is entangled with the state
of the de-excited spin. This process can be performed many times to
generate a large cluster of entangled photons.

Schwartz and coworkers\citep{Schwartz2016} used the QD-confined dark-exciton
(DE) as entangler. The DE has an integer total spin (2) and the short-range
exchange interaction between its electron and hole removes its spin
degeneracy \citep{Bayer2002}. As a result, no external field is required
for the DE to precess and be used as an entangler\citep{Schwartz2016}.

There are, however, advantages for using the electron or even better,
the heavy hole, which has comparable spin coherence time to that of
the DE \citep{Cogan2018}, as an entangler. Single carriers as opposed
to the DE, have half integer total spin (1/2 and 3/2, respectively)
and therefore in the absence of external field are Kramers' degenerate.
Therefore, an external field is required for them to precess, and
its magnitude can be utilized to accurately tune the precession period.
This allows optimization of the entanglement robustness and the generation
rate of the entangled multiphotons \citep{Schwartz2016}. Moreover,
in the single carrier case the emitted photons are indistinguishable
\citep{Kiraz2004,Gantz2017}. This is true as long as the the external
field-induced Zeeman-splitting of the spin states are smaller than
the spectral width of the emitted photons. In this particular regime,
the Zeeman interaction is comparable in magnitude to the hyperfine
interaction induced by the surrounding nuclei spins. Thereby, one
must consider both interactions in designing QD based devices as entangled
light sources.

Studies of the central-spin evolution under the influence of the hyperfine
and quadrupole interactions with the nuclei were recently published
\citep{Bechtold2015,Cogan2018}. Few distinct temporal domains in
the central spin evolution were observed. During the first temporal
domain, which is relevant to this work, the Overhauser field, due
to frozen fluctuations in the nuclei spins, can be considered static
and the central spin precesses about this field direction.

In this work, we study experimentally and theoretically the central-spin
evolution and decoherence mechanism of both the QD confined electron
\citep{Bluhm2010,Braun2005} and heavy-hole (HH) \citep{Gerardot2008,Brunner2009,Eble2009,Fras2011,De_Greve_2011,Li2012,Prechtel2016}
spins. We vary the ratio between the magnetically induced Zeeman interaction
and the hyperfine interaction with the central spin and measure the
temporal evolution of the spin purity.

In addition to its scientific importance, this work provides tools
for state tomography, engineering and optimizing QDs based devices
for generating photonic cluster and graph states. These multiphoton
entangled states are important resources for quantum repeaters and
entanglement distribution \citep{Zwerger2012,Munro2012,Azuma2015,Buterakos2017},
enabling implementations of quantum information communication protocols.

\begin{figure}
\begin{centering}
\includegraphics[width=1\columnwidth]{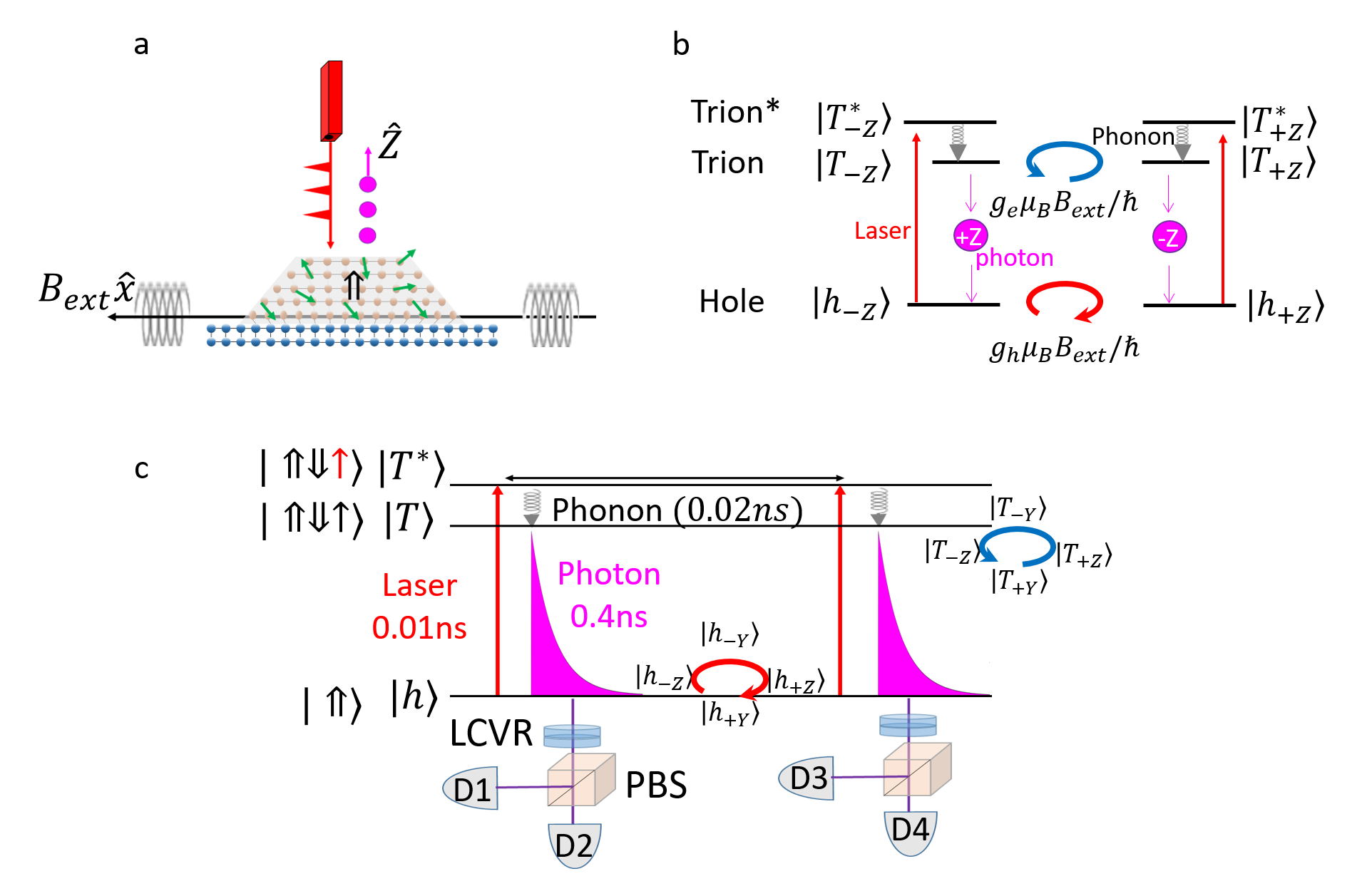}
\par\end{centering}
\caption{\label{fig:methods}a) The QD sample. 12~ps laser $\pi$-pulses (marked
in red) resonantly excite the central spin. The excited spin then
returns to its ground state by emitting single photons (pink circles).
The spin evolves under the joint influence of the external and nuclear
magnetic fields. Green arrows represent the randomly distributed nuclear
spins. b) The polarization selection rules for the hole-Trion optical
transition. $\pm Z$ is the polarization of the exciting laser and
the emitted photons. c) The experimental setup for initializing and
probing the central spin evolution. Liquid-crystal-variable-retarders
(LCVR's) and polarizing beam-splitters (PBSs) are used to project
the photons polarization. By selecting the polarization of the exciting
pulses and that of the projected photons, while setting the time difference
between the pulses, one can fully investigate the electron and hole
spins evolution.}
\end{figure}
\begin{figure}
\begin{centering}
\includegraphics[width=1\columnwidth]{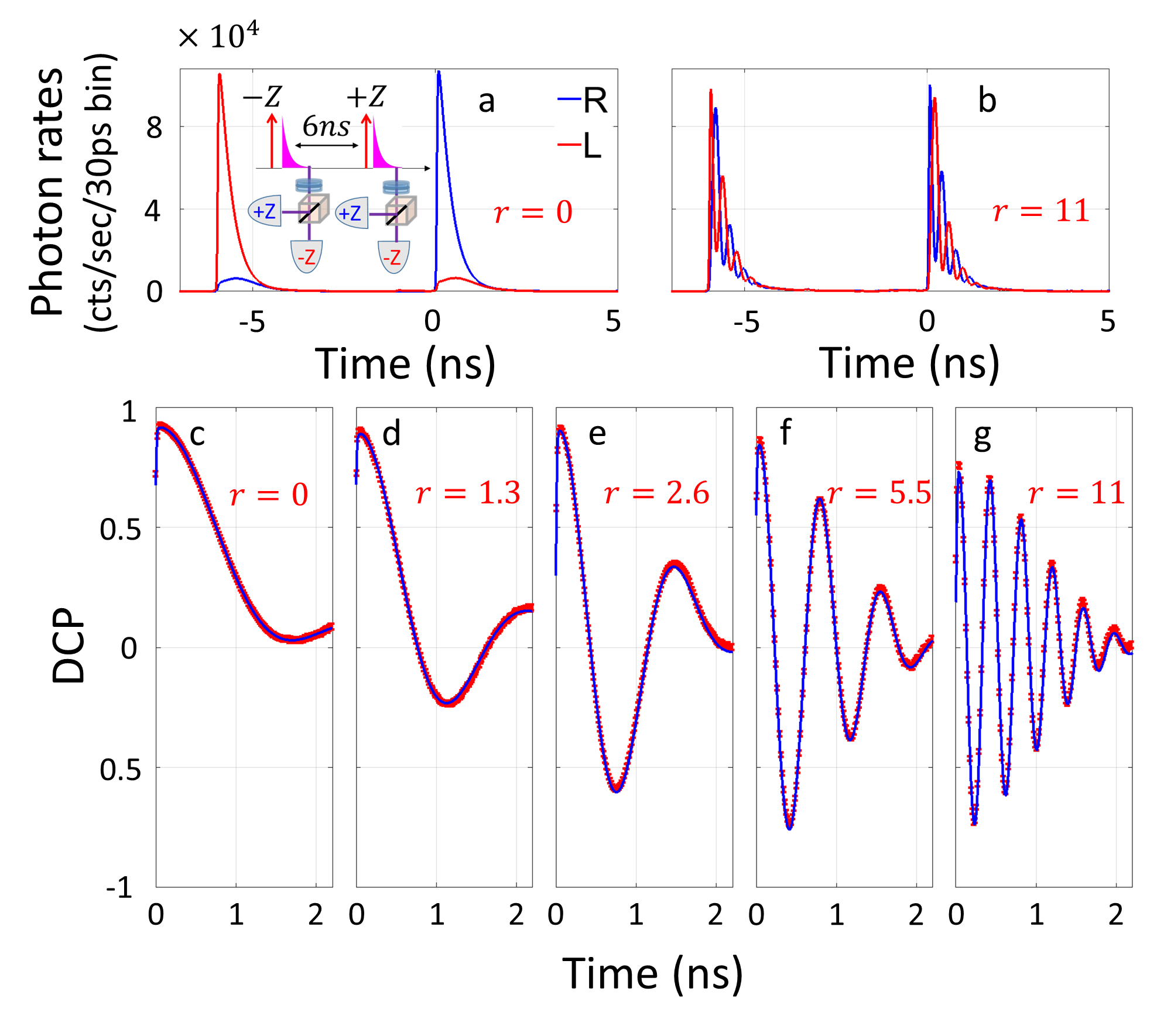}
\par\end{centering}
\caption{\label{fig:Trion}Measurements of the Trion spin-evolution. Polarization
sensitive time-resolved PL emission from the positive trion transition
for a Zeeman-Hyperfine ratio (Eq.~\ref{eq:ratio parameter}) of a)
$r=0$, and b) $r=11$. The inset in a) describes the pulse (red arrows)
and detection (magenta markings) sequence used for these measurements.
The trion spin is alternately initialized to spin down and up to prevent
buildup of Overhauser-field in the sample. c) - g) Time resolved degree
of circular polarization ($D_{cp}$, Eq.~\ref{eq:DCP}) of the PL
emission for various r parameters. Red marks represent the measured
data and the overlaid blue lines represent our central-spin model
best fits (see SM). From these multiple fits, we deduce the electron's
g-factor $g_{e}$, the interaction energy with the nuclear environment
$\gamma_{e}$, and the Trion's radiative time $\tau_{photon}$, as
summarized in Table.~\ref{tab:Device-parameters}.}
\end{figure}

The primary decoherence mechanism of electronic spin qubits in QDs
is the hyperfine-interaction with the $\sim10^{5}$ nuclei spins in
the QD \citep{Gammon_2001,Efros2002,Khaetskii2002,Fischer2008}. The
Zeeman interaction with the external field does not cause decoherence
but induces coherent precession around the field direction. The two
interactions can be described by a Hamiltonian 
\begin{equation}
H=\frac{1}{2}\vec{C}\cdot\hat{\sigma}\label{eq:Hamiltonian_general}
\end{equation}
 acting on the electronic spin, with $\hat{\sigma}$ being the Pauli
matrices describing the spin. $\vec{C}=\overline{g}\mu_{B}\vec{B}_{ext}+\overline{\gamma}\vec{B}_{N}$
is the joint Zeeman and hyperfine interactions, where $\overline{g}$
and $\overline{\gamma}$ are the Lande-factor and hyperfine interaction
tensors, $\vec{B}_{ext}$ and $\vec{B}_{N}$ are the external and
Overhauser field. The Overhauser field is treated as random, having
Gaussian distribution with zero mean and a variance $\sigma$ (not
to be confused with the Pauli matrices). Defining a modified unitless
magnetic field $\vec{B}=\vec{B}_{N}/\sigma$ with root mean square
of one and redefining the tensor $\overline{\gamma}$ having energy
units, the corresponding probability distribution then takes the standard
form $dP(B)=\frac{1}{(2\pi)^{3/2}}\exp\left(-\frac{1}{2}B^{2}\right)d^{3}B$.

While the conduction-electron interaction with the nuclei, which we
denote by $\gamma_{e}$, is mostly isotropic \citep{Efros2002}, the
interaction of the valence heavy-hole for which the orbital momentum
and the spin are aligned with the growth direction is anisotropic
\citep{Fischer2008}. We denote by $\gamma_{h_{z}}$ ($\gamma_{h_{p}}$)
the interaction of the valence heavy-hole spin with the Overhauser
field component along (perpendicular to) the growth axis $\hat{z}$
\citep{Cogan2018}. Defining the externally applied magnetic field
$B_{ext}$ as the x-direction, and assuming that the tensors g and
$\gamma$ are diagonal (see SM) the vector C is given by: 
\[
\vec{C}=[g_{x}\mu_{B}B_{ext}+\gamma_{x}B_{x},\gamma_{y}B_{y},\gamma_{z}B_{z}].
\]

and we use it to derive expressions for the temporal evolution of
both the heavy-hole and electron spins (see SM). We then investigate
the spin evolution of both carriers for various external field strengths,
affecting the Zeeman to Hyperfine interaction ratio: 
\begin{equation}
r=C_{zeeman}/C_{hf}^{\perp}=\frac{g_{x}\mu_{B}B_{ext}}{\sqrt{\gamma_{y}^{2}+\gamma_{z}^{2}}}.\label{eq:ratio parameter}
\end{equation}
.

\begin{figure*}
\begin{centering}
\includegraphics[width=1\textwidth]{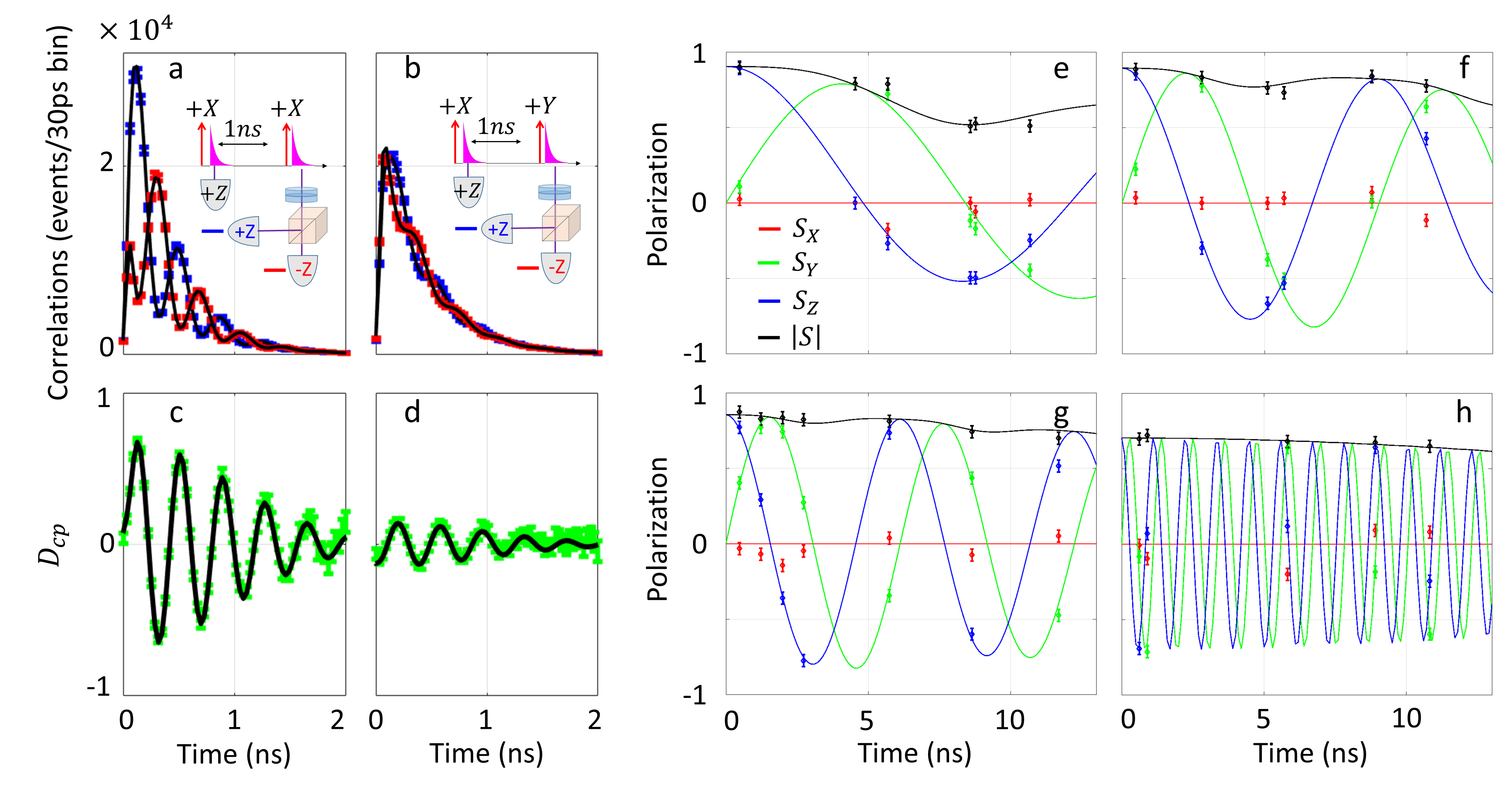}
\par\end{centering}
\caption{\label{fig:Hole} The heavy-hole spin evolution. a)-d) Full tomography
of the HH spin \citep{Cogan2020} at time t=1ns after its initialization
to +Z by detecting the first circularly polarized emitted photon,
for $r=30$ (Eq.~\ref{eq:ratio parameter}). a) and b) are polarization-sensitive
time-resolved intensity correlation measurements of the Trion emission
after the second pulse, where blue (red) marks represent measurements
in which the two detected photons are co- (cross-) circularly polarized.
Black lines describe the best fitted model (see SM). The insets to
a) and b) describe the polarized pulse sequence (red arrows) and the
resulted emission detection (magenta). c) and d) describe the degree
of circular polarization ($D_{cp}$, see Eq.~\ref{eq:DCP}) deduced
from the measurements in a) and b), respectively. The HH-spin state
$[S_{x},S_{y},S_{z}]$, at the second pulse time is extracted from
the fits in a), b), c) and d) \citep{Cogan2020}. Similar spin tomography
measurements are performed for various pulse separation times (t)
and for different ratios r .The HH spin-evolution is presented in
e) $r=1.8$, in f) $r=3.6$, in g) $r=5.4$, and in h) $r=30$. The
red, green, and blue marks describe the HH projections $S_{x}(t)$,
$S_{y}(t)$, and $S_{z}(t)$, respectively, while the black marks
describe the spin-purity $|S|=\sqrt{S_{x}^{2}+S_{y}^{2}+S_{z}^{2}}$.
The bars represent the measurement uncertainties and the color matched
solid lines represent model-fit to the measured data (see SM). From
these fits, we extract the HH in-plane g-factor $g_{h}$, and the
hyperfine-interaction energies $\gamma_{h_{p}}$ and $\gamma_{h_{z}}$
(Table.~\ref{tab:Device-parameters}).}
\end{figure*}

\begin{figure}
\centering{}\includegraphics[width=1\columnwidth]{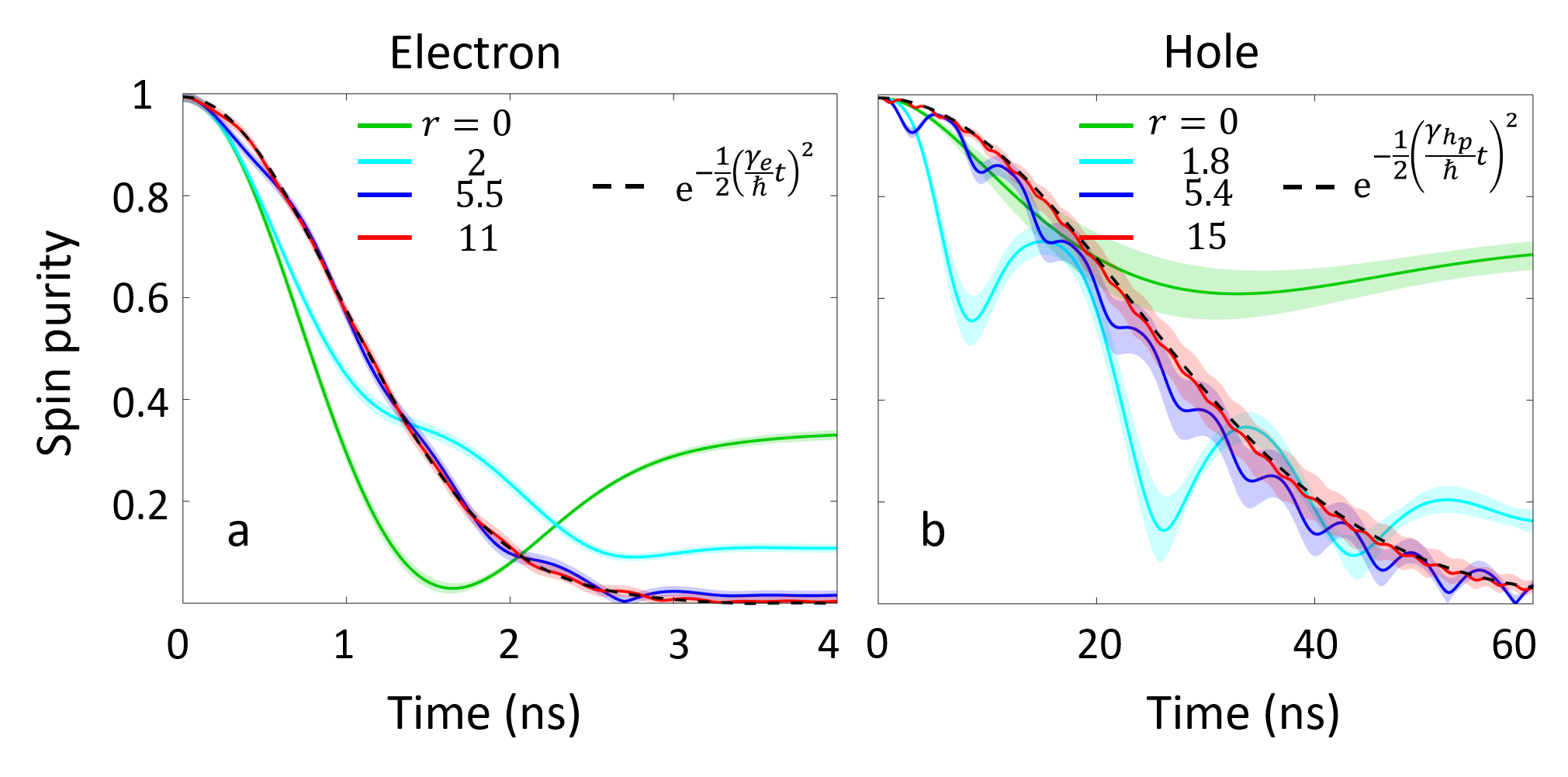}\caption{\label{fig:Spin_decoherence} Model calculated central spin purity
$\left|S(t)\right|=\sqrt{S_{x}^{2}(t)+S_{y}^{2}(t)+S_{z}^{2}(t)}$
vs. time for a) the electron, and b) the heavy-hole. The spin purity
is calculated using the central spin model (see SM and Table.~\ref{tab:Device-parameters}).
The spin is initialized in the z-direction ($S_{z}(0)=1$) and the
external field is applied in the x-direction. The colored lines represent
the purity vs. time from initialization for various Zeeman-Hyperfine
ratios r (Eq.~\ref{eq:ratio parameter}). The shaded areas represent
the uncertainties in the measured values of Table.~\ref{tab:Device-parameters}.
The calculations result from Fig.~\ref{fig:Trion} for the electron
and Fig.~\ref{fig:Hole} for the hole. The dash black line represents
the analytic solution $\left|S(t)\right|=e^{-\frac{1}{2}\left(\frac{\gamma_{x}}{\hbar}t\right)^{2}}$
describing the purity in the limit $r>>1.$}
\end{figure}

We mark the HH spin along the shortest axis of the QD's 3-D potential
trap (marked +Z in Fig.~\ref{fig:methods}a) by $\ket{\Uparrow}$
($\ket{\Downarrow})$ for spin up (down) state. Resonantly tuned laser
pulse photogenerates an extra electron-hole pair in the QD, converting
the HH to a positive Trion ($\Uparrow\Downarrow\uparrow$). The ground
level of the positive trion is composed of two paired HHs with opposite
spins and a single conduction band electron. The selection rules for
optical transitions between the HH spin states and that of the trion
form a $\pi$-system \citep{Cogan2018} with the following optical
transitions: 
\begin{align}
\ket{\Uparrow} & \stackrel{\ket{-Z}}{\longleftrightarrow}\ket{\Uparrow\Downarrow\uparrow},\nonumber \\
\ket{\Downarrow} & \stackrel{\ket{+Z}}{\longleftrightarrow}\ket{\Downarrow\Uparrow\downarrow}.\label{eq:Selection_rules}
\end{align}

where $\ket{+Z}$ and $\ket{-Z}$ denote the right- and left-hand
circular polarization states of the photon inducing the optical transition
between the ground level qubit (HH) and the excited qubit (Trion)
, as described in Fig.~\ref{fig:methods}b.

We note that this system allows us to investigate both the valence
band HH and the conduction band electron as central spins. The HH
is investigated through the ground level qubit while the electron
through the excited level trion. The magnetic interaction (Eq.~\ref{eq:Hamiltonian_general})
for the paired HH spins vanishes, leaving only the electron to consider
\citep{Cogan2018}.

Fig.~\ref{fig:methods}a and \ref{fig:methods}b show a schematic
description of the InAs in GaAs self-assembled QD sample and the polarization
selection rules for the hole-Trion optical transition, respectively.
The externally applied in-plane magnetic field induces coherent evolution
of the central spins, which precess between their spin-up and spin-down
states. The interaction of the electronic spin with the nuclear spin
environment inflicts decoherence on the evolving central spins.

We use two-laser-pulses to initialize and to probe the HH and the
trion qubits as described in Fig.~\ref{fig:methods}c. Each laser
pulse excites the HH to an excited-positive-trion level, where the
unpaired electron is in its respective second energy level. The excited
electron decays to its ground level forming a ground level trion by
emitting an optical phonon within about $\sim20ps$. The trion then
decays radiatively by emitting a photon within about $400ps$. We
project the photon's polarization using pairs of liquid-crystal-variable-retarders
(LCVRs) followed by polarizing beam-splitters (PBS), transmission
gratings for spectral filtering and superconducting single-photon
counters for detection. The overall system efficiency is about 1\%
and its temporal resolution is about $30ps$. We initialize the HH
and trion spin qubits by controlling the polarization of the resonant
laser pulses. At the same time we also probe the state of these qubits
by measuring the polarization of the emitted photons.
\begin{table}
\caption{\label{tab:Device-parameters}The QD g-tensor and hyperfine-tensor
components. Here $\gamma_{h_{p}}$, $\gamma_{h_{z}}$, and $\gamma_{e}$
are the interaction energies of the hole and electron spins with the
nuclear spin environment and $g_{h}^{x}$ and $g_{e}^{x}$ are the
relevant HH and electron g-tensor components. The positive Trion radiative
time is $\tau_{photon}$.}

\centering{}%
\begin{tabular}{|c|c|c|}
\hline 
 & This work & Literature\tabularnewline
\hline 
\hline 
$\gamma_{h_{p}}$ & $0.029\pm0.005\mu eV$ & $0.031$\citep{Cogan2018}, $0.047$\citep{Eble2009}\tabularnewline
\hline 
$\gamma_{_{h_{z}}}$ & $0.12\pm0.02\mu eV$ & $0.11$\citep{Cogan2018}, $0.081$\citep{Eble2009}\tabularnewline
\hline 
$\gamma_{e}$ & $0.693\pm0.006\mu eV$ & $0.34$\citep{Cogan2018}, $0.33$\citep{Bechtold2015}\tabularnewline
\hline 
$g_{h}$ & $-0.128\pm0.002$ & \tabularnewline
\hline 
$g_{e}$ & $0.367\pm0.004$ & \tabularnewline
\hline 
$\tau_{photon}$ & $0.398\pm0.004ns$ & \tabularnewline
\hline 
\end{tabular}
\end{table}
 \\

Fig.~\ref{fig:Trion} describes how we measure the spin evolution
of the positive trion. We initialize the trion qubit to spin-up and
spin-down states alternately using -Z and +Z polarized 12~ps long
laser $\pi$-pulses. This is done in order to avoid the accumulation
of unwanted Overhauser-field \citep{Knight1949,Overhauser1953,Barnes2011,Carter2014,Cogan2018}.
We then monitor the spin evolution during the optical recombination
by projecting the emitted photon on -Z and +Z polarization basis.
Fig.~\ref{fig:Trion}a - \ref{fig:Trion}b show the time-resolved
PL emission of the emitted photons for $r=0$ and $r=11$, respectively.
Since the trion in the $\ket{T_{+Z}}$ state emits -Z photon while
$\ket{T_{-Z}}$ state emits +Z photon, the spin evolution of the trion
can be deduced from the polarization sensitive time resolved PL. For
the Zeeman-Hyperfine ratio $r=11$ case (Eq.~\ref{eq:ratio parameter}),
the trion's spin-precession around the external magnetic field is
clearly visible, while for the $r=0$ case, one can only observe the
exponential radiative decay. To increase this measurement's sensitivity,
we look at the degree of circular polarization ( $D_{cp}$) of the
emission defined by 
\begin{equation}
D_{cp}(t)=\frac{I_{+Z}(t)-I_{-Z}(t)}{I_{+Z}(t)+I_{-Z}(t)},\label{eq:DCP}
\end{equation}
where $I_{+Z}(t)$ ($I_{-Z}(t)$) denotes the measured polarized PL
intensity. The $D_{cp}$(t) distills the information on the spin evolution
from the characteristic exponential radiative decay \citep{Cogan2018,Cogan2020}
. Using Eq.~\ref{eq:Selection_rules}, it is straightforward to show
that the $D_{cp}(t)$ measures the trion's spin projection on Z-axis
- $S_{z}(t)$.

Fig.~\ref{fig:Trion}c-g shows the measured $D_{cp}(t)$ for five
different r cases. For $r=0$, the spin polarization evolves only
due to the Hyperfine interaction. It first decays within \textasciitilde 1.65ns
and then revives back to 1/3 of its initial value \citep{Efros2002,Bechtold2015,Cogan2018}.
For $r>0$, the spin evolves around the vector sum of $\vec{B}_{ext}$
and $\vec{B}_{N}$. One notes that stronger external fields (higher
r) result in a decrease of the initial spin polarization ($S_{z}(0)$=
0.92 for r=0, $S_{z}(0)$= 0.71 for r=11). This decrease is due to
the detectors' finite temporal resolution (\textasciitilde 30~ps).

In Fig.~\ref{fig:Hole}e-h we similarly present the measured HH-spin
temporal evolution for four different r ratios. The hole state is
initialized to -Z state by projecting the first photon on +Z polarization.
For measuring the HH spin state as a function of time after this initialization
we utilize a measurement technique which provides a full tomography
of the HH spin \citep{Cogan2020}. Fig.~\ref{fig:Hole}a-d demonstrates
the application of this technique 1ns after initialization. The evolving
hole state is promoted to the trion state by a second 1ns-delayed
$\pi$-pulse, linearly polarized +X (Horizontal), and +Y (Diagonal).
+X polarized pulse promotes an arbitrary hole spin state $\alpha\ket{h_{+Z}}+\beta\ket{h_{-Z}}$
to the trion state: $\alpha\ket{T_{+Z}}+\beta\ket{T_{-Z}}$, while
+Y polarized pulse results in a $\alpha\ket{T_{+Z}}+i\beta\ket{T_{-Z}}$
trion. The $D_{cp}$ of the emission as a function of time for both
cases provide the means for a full tomography of the HH spin state
at the second excitation pulse \citep{Cogan2020}. The tomography
or the HH spin projections $S_{x}(t)$, $S_{y}(t)$, and $S_{z}(t)$,
for t=1ns, can be quite faithfully extracted from the best fitted
model calculations to the measured points in Figs. \ref{fig:Hole}a-d.
These fits are presented by the solid black lines overlaid on the
experimental measurements.

In Fig.~\ref{fig:Hole}e-h, we present by red, green, and blue, the
measured $S_{x}(t)$ , $S_{y}(t)$, and $S_{z}(t)$, respectively,
for various r ratios. The black points represent the spin purity which
we define as $\left|S\right|=\sqrt{S_{x}^{2}+S_{y}^{2}+S_{z}^{2}}$.
The color matched solid lines represent the calculated values (see
SM) using the parameters in Table.~\ref{tab:Device-parameters}

In Fig.~\ref{fig:Spin_decoherence} we present the model-calculated
time-resolved purity of the electron (in a) and HH (in b) spins, for
various Zeeman-Hyperfine ratios $r$ (see Eq.~\ref{eq:ratio parameter}).
The shaded areas represent one standard deviation of the model's uncertainties
in the measured Table.~\ref{tab:Device-parameters} values. For $r=0$,
the spin depolarizes, reaches minimum and then partially revives.
Spin precession around the frozen fluctuation of the nuclear field
adequately describes this observation \citep{Efros2002,Bechtold2015,Cogan2018}.
Since we initialize the spin in the z-direction, $\gamma_{x}$ and
$\gamma_{y}$ are the only relevant hyperfine tensor components contributing
to the central spin depolarization. $\gamma_{z}$, in contrast, pins
the spin to its initial direction. Consequently, the spin reaches
minimum within $t_{min}=h/\left(2(\gamma_{x}+\gamma_{y})\right)$,
with h being the plank constant. This time is given by the hyperfine-induced
spin precession perpendicular to the initialization direction. In
addition, the value that the purity reaches after the revival depends
on the ratio between $\gamma_{z}$ and $\gamma_{x}+\gamma_{y}+\gamma_{z}$.
Since the hyperfine tensor is isotropic for the electron its purity
revives to 1/3. On the other hand, the anisotropic HH revives to \textasciitilde 2/3.

For $r>1$, as the external field increases, the Zeeman interaction
becomes greater than the hyperfine interaction. The Zeeman induced
coherent spin precession averages the influence of the hyperfine interaction
perpendicular to the field, effectively reducing $\gamma_{y}$ and
$\gamma_{z}$. This effect can be viewed as a natural dynamical decoupling.
This in turn results in temporal oscillations of the central spin
purity. The frequency of these oscillations increases linearly with
$r$ while their amplitude decays. In addition, the spin revival peak
reduces with the field as the effective ratio $\gamma_{z}/(\gamma_{x}+\gamma_{y}+\gamma_{z})$
decreases. In the limit $r>>1$, the temporal dependence of the spin
purity decay can be expressed analytically as $\left|S(t)\right|=e^{-\frac{1}{2}\left(\frac{\gamma_{x}}{\hbar}t\right)^{2}}$
depending only on the hyperfine interaction parallel to the field
direction (see SM).\\

In summary, we present a comprehensive theoretical study of a central
spin evolution interacting with nuclear spins in its vicinity, in
the presence of externally applied magnetic field. Our studies are
compared with comprehensive experimental studies of the evolution
of the electron and heavy-hole spins confined in the same semiconductor
quantum dot. We show that the central spin model well-describes the
measured spins evolution. For large external field the central spin
coherence decays like a Gaussian depending only on the hyperfine interaction
parallel to the field's direction. In lower fields, in which the Zeeman
interaction is comparable in magnitude to the hyperfine interaction
with the nuclear spins, the spin purity oscillates in time during
its dephasing. Under these conditions the Zeeman interaction is also
smaller than the radiative linewidth. This is essential for entangled
light sources. Our comprehensive study is therefore an important step
towards bringing quantum dots based indistinguishable and entangled
photon sources closer to real applications.\\

The support of the Israeli Science Foundation (ISF), and that of the
European Research Council (ERC) under the European Union\textquoteright s
Horizon 2020 research and innovation programme (Grant Agreement No.
695188) are gratefully acknowledged.

\bibliographystyle{aipnum4-1}

\end{document}


\title{Supplemental material for ``The coherence of quantum dot confined
electron- and hole-spin in low external magnetic field''}
\author{Dan Cogan}
\affiliation{The Physics Department and the Solid State Institute, Technion\textendash Israel
Institute of Technology, 3200003 Haifa, Israel}
\author{Zu-En Su}
\affiliation{The Physics Department and the Solid State Institute, Technion\textendash Israel
Institute of Technology, 3200003 Haifa, Israel}
\author{Oded Kenneth}
\affiliation{The Physics Department and the Solid State Institute, Technion\textendash Israel
Institute of Technology, 3200003 Haifa, Israel}
\author{David Gershoni}
\email{dg@physics.technion.ac.il}

\affiliation{The Physics Department and the Solid State Institute, Technion\textendash Israel
Institute of Technology, 3200003 Haifa, Israel}

\maketitle
\global\long\def\ket#1{\left|#1\right\rangle }%
\global\long\def\im{\operatorname{Im}}%
\global\long\def\bra#1{\left\langle #1\right|}%

We describe the temporal evolution of the polarization of a QD confined
central electronic spin in the presence of an externally applied magnetic
field and the influence of the nuclear spins, which comprise the QD
and its vicinity. As the central-spin, we consider either an electron
or a heavy hole.

As both cases involve a two-level spin system (a qubit), they may
be described using the Pauli matrices $\sigma_{x},\sigma_{y},\sigma_{z}$,
and the effective Hamiltonian must take the form 
\begin{equation}
H=\frac{1}{2}\vec{C}\cdot\hat{\sigma},
\end{equation}
for some vector $\vec{C}=\vec{C}^{hf}+\vec{C}^{Zeeman}$ describing
the spin's hyperfine interaction with the QD nuclei and the Zeeman
interaction with the external magnetic field. The exact expression
of $\vec{C}$ is different, of course, for each type of central spin.\\
The hyperfine interaction between the central spin and the effective
magnetic (Overhauser) field generated by the \textasciitilde 10\textasciicircum 5
nuclear spins in the QD is defined by 
\[
\vec{C}^{hf}=\overline{g}\mu_{B}\vec{B}_{N},
\]

where $\overline{g}$ and $\mu_{B}$ are the central spin's lande-factor
tensor and Bohr magneton, and $\vec{B}_{N}$ is the Overhauser field.
The Overhauser field depends on the nature of the interaction between
the spin and the nuclear environment \citep{Efros2002,Eble2009,Cogan2018}.

Assuming that different nuclear spins are not correlated allows one
to treat $\vec{B}_{N}(t)$ as having isotropic Gaussian random distribution
satisfying 
\begin{equation}
\langle\vec{B_{N}}\rangle=0,\quad\langle B_{Nx}^{2}\rangle=\langle B_{Ny}^{2}\rangle=\langle B_{Nz}^{2}\rangle=\sigma^{2},
\end{equation}

where $\sigma$ is the width of the Gaussian distribution \citep{Efros2002}
(not to be confused with the Pauli matrices).

It is then convenient to define a modified unitless magnetic field
$\vec{B}=\vec{B}_{N}/\sigma$ with root mean square of one and absorb
$\sigma$ into a redefined tensor $\overline{\gamma}$ having energy
units. $H^{hf}=\frac{1}{2}\vec{C}^{hf}\cdot\vec{\sigma}$ then express
the spin's hyperfine interaction with $\vec{C}_{e}^{hf}=\overline{\gamma}\vec{B}$
where $\overline{\gamma}=\overline{g}\mu_{B}\sigma$ is the spin hyperfine
coupling tensor.

While for the conduction electron, which has s-wave molecular symmetry
the hyperfine tensor is approximately a scalar: \citep{Efros2002}
\[
\overline{\gamma}_{e}=\left(\begin{array}{ccc}
\gamma_{e}\\
 & \gamma_{e}\\
 &  & \gamma_{e}
\end{array}\right),
\]
for the heavy hole, this tensor is approximated by a diagonal but
anisotropic tensor \citep{Eble2009}
\[
\overline{\gamma}_{h}=\left(\begin{array}{ccc}
\gamma_{h_{p}}\\
 & \gamma_{h_{p}}\\
 &  & \gamma_{h_{z}}
\end{array}\right),
\]
where the in-plane coupling component $\gamma_{h_{p}}$ does not strictly
vanish for the heavy-hole due its mixing with the light-hole \citep{Eble2009}.
Thus $\gamma_{h_{z}}>\gamma_{h_{p}}$.

Strictly speaking, the normalized Overhauser $\vec{B}$ field that
the hole feels maybe different than the field that the electron does.
This is due to differences in the interaction of the electron and
the hole spins with the nuclei. For our purpose, however, it is sufficient
that the fields have the same Gaussian statistics. For the moment,
we allow the functional relation between $\vec{C}$ and $\vec{B}$
to be arbitrary, and since our discussion is independent of these
relations, it applies to both cases.

The Zeeman interaction between the central spin and an externally
applied magnetic field $\vec{B}_{ext}$, is given by

\begin{equation}
\vec{C}^{zeeman}=\overline{g}\mu_{B}\vec{B}_{ext}.
\end{equation}

It is fairly customary to assume that $\overline{g}$, the Lande-factor
tensor describing the Zeeman interaction, is diagonal in our QD for
both the electron and the hole, and taking the external magnetic field
along the x-direction, the total interaction then assumes the form
of a vector, given by

\begin{align}
\left[C_{x},C_{y},C_{z}\right]^{(electron)} & =[g_{e}\mu_{B}B_{ext}+\gamma_{e}B_{x},\gamma_{e}B_{y},\gamma_{e}B_{z}],\label{eq:interaction_electron-1-1}\\
\left[C_{x},C_{y},C_{z}\right]^{(hole)} & =[g_{h}\mu_{B}B_{ext}+\gamma_{h_{p}}B_{x},\gamma_{h_{p}}B_{y},\gamma_{h_{z}}B_{z}],\label{eq:interaction_hole-1-1}
\end{align}

for the electron and hole, respectively. In the following we use $g_{e}$,
$g_{h}$, $\gamma_{e}$, $\gamma_{h_{p}}$, and $\gamma_{h_{z}}$
as fitting parameters, to best describe all the experimental observations.

$\vec{B}$ and $\vec{C}$ can be treated as time-independent at short
times, and one readily finds the solution 2b
\begin{align}
\vec{S}(t) & =\frac{\vec{S}_{0}\cdot\vec{C}}{C^{2}}\vec{C}+\left(\vec{S}_{0}-\frac{\vec{S}_{0}\cdot\vec{C}}{C^{2}}\vec{C}\right)\cos\left(\frac{C}{\hbar}t\right)\label{eq. Spin-evolution}\\
 & -\frac{\vec{S}_{0}\times\vec{C}}{C}\sin\left(\frac{C}{\hbar}t\right),\nonumber 
\end{align}
where $\vec{S}_{0}=\vec{S}(0)$ is the central spin initial value.
One may rewrite this relation in the form $\overrightarrow{S}(t)=\bar{G}(t)\overrightarrow{S}_{0}$
where $\bar{G}(t)$ is a 3 \texttimes{} 3 tensor whose elements can
be easily read of Eq.~\ref{eq. Spin-evolution}.

The above relation may be interpreted either as a relation between
(Heisenberg picture) operators or as a relation between their expectation
values. In the following, it will be more convenient to take the latter
point of view. (Note that the expectation value $\left\langle \vec{S}\right\rangle $
contains complete information about the quantum state for a two-state
system.)

The probability distribution of B takes the standard form 
\begin{equation}
dP=\frac{1}{(2\pi)^{3/2}}\exp\left(-\frac{1}{2}B^{2}\right)d^{3}B.\label{p1-1-1}
\end{equation}

So the averaged spin is given by 
\[
\left\langle \overrightarrow{S}(t)\right\rangle =\left[\int\bar{G(t)}dP\right]\overrightarrow{S}_{0}=\bar{\mathbf{G(t)}}\overrightarrow{S}_{0},
\]
where $\bar{\mathbf{G(t)}}=\int\bar{G(t)}dP$.

Under these assumptions, one may employ symmetries under reflections
of the Overhauser field to show that the 3 \texttimes{} 3 tensor \textbf{G(t)}
simplifies (basically due to the vanishing of odd integrals) into
the form

\begin{equation}
\bar{\mathbf{G(t)}}=\left[\begin{array}{ccc}
\mathbf{G}_{xx}(t)\\
 & \mathbf{G}_{yy}(t) & \mathbf{G}_{yz}(t)\\
 & -\mathbf{G}_{yz}(t) & \mathbf{G}_{zz}(t)
\end{array}\right]
\end{equation}
with
\begin{align}
\mathbf{G}_{xx}(t)=\int\frac{1}{C^{2}} & \left[C_{x}^{2}+\left(C_{y}^{2}+C_{z}^{2}\right)\cos\left(\frac{C}{\hbar}t\right)\right]dP,\nonumber \\
\mathbf{G}_{yy}(t)=\int\frac{1}{C^{2}} & \left[C_{y}^{2}+\left(C_{x}^{2}+C_{z}^{2}\right)\cos\left(\frac{C}{\hbar}t\right)\right]dP,\nonumber \\
\mathbf{G}_{zz}(t)=\int\frac{1}{C^{2}} & \left[C_{z}^{2}+\left(C_{x}^{2}+C_{y}^{2}\right)\cos\left(\frac{C}{\hbar}t\right)\right]dP,\nonumber \\
\mathbf{G}_{yz}(t)= & -\int\frac{C_{x}}{C}\sin\left(\frac{C}{\hbar}t\right)dP.\label{eq:G_elements-1}
\end{align}

We solve the integrals in Eq.~\ref{eq:G_elements-1} numerically
with the parameters of Table. 1. We compare the results of the spin
evolution measurements in Fig. 2 for the electron and Fig. 3 for the
hole with our model for several Zeeman-hyperfine interaction ratio's
r defined by
\begin{equation}
r=C_{zeeman}/C_{hf}^{\perp}=\frac{g_{x}\mu_{B}B_{ext}}{\sqrt{\gamma_{y}^{2}+\gamma_{z}^{2}}}.\label{eq:ratio parameter}
\end{equation}

There is a good agreement between the model and the measured results.

The PL measurements in Figs. 2a-2b and Figs. 3a-3b are fitted with
\citep{Cogan2020} 
\begin{align*}
I_{+Z}(t) & =I(0)\exp(-t/\tau_{photon})\left(1-S_{z}(t)\right)\\
I_{-Z}(t) & =I(0)\exp(-t/\tau_{photon})\left(1+S_{z}(t)\right),
\end{align*}

where $\tau_{photon}$ is the radiative time of the Trion-HH optical
transition, $I(0)$ is the initial emission intensity, and $I_{+Z}(t)$
($I_{-Z}(t)$) denotes the measured PL intensity projected on right-hand
(left-hand) circular polarization. $S_{z}(t)$ is the z component
of the trion spin at time t, calculated numerically using Eq.~\ref{eq:G_elements-1}
with the values of Table. 1 for the electron. There is a good agreement
between the model and the measured results.

Next, we expand Eq.~\ref{eq:G_elements-1} in the limit of high external
magnetic field $B_{ext}\gg\gamma/(g\mu_{B})$ or r>\textcompwordmark >1.
In this limit, The tensor $\bar{\mathbf{G}(t)}$ is given by

\begin{align}
\mathbf{G}_{xx}(t) & =1-\frac{1}{r^{2}}\left(1-\cos\left(\omega_{0}t\right)e^{-\frac{1}{2}\left(\frac{t}{T_{2}}\right)^{2}}\right)\nonumber \\
 & +O\left(B_{ext}^{-3}\right)\nonumber \\
\mathbf{G}_{yy}(t) & =e^{-\frac{1}{2}\left(\frac{t}{T_{2}}\right)^{2}}\cos\left(\omega_{0}t\right)\nonumber \\
 & -e^{-\frac{1}{2}\left(\frac{t}{T_{2}}\right)^{2}}\frac{\omega_{0}}{r^{2}}t\sin\left(\omega_{0}t\right)+O\left(B_{ext}^{-2}\right)\nonumber \\
\mathbf{G}_{zz} & (t)=e^{-\frac{1}{2}\left(\frac{t}{T_{2}}\right)^{2}}\cos\left(\omega_{0}t\right)\nonumber \\
 & -e^{-\frac{1}{2}\left(\frac{t}{T_{2}}\right)^{2}}\frac{\omega_{0}}{r^{2}}t\sin\left(\omega_{0}t\right)+O\left(B_{ext}^{-2}\right)\nonumber \\
\mathbf{G}_{yz} & (t)=-e^{-\frac{1}{2}\left(\frac{t}{T_{2}}\right)^{2}}\sin\left(\omega_{0}t\right)\nonumber \\
 & -e^{-\frac{1}{2}\left(\frac{t}{T_{2}}\right)^{2}}\frac{\omega_{0}}{r^{2}}t\cos\left(\omega_{0}t\right)+O\left(B_{ext}^{-2}\right),\label{eq:G high field}
\end{align}
where $\omega_{0}=\frac{g\mu_{B}B_{ext}}{\hbar}$ is the Zeeman frequency,
r is the Zeeman-hyperfine ratio parameter, defined in Eq.~\ref{eq:ratio parameter},
and $T_{2}$ is the spin decay time.

In this large external field limit the total effective field $\vec{C}$
is almost parallel to the x-axis, and therefore the corresponding
spin component $S_{x}$ remains almost unchanged (i.e. $G_{xx}\simeq1$)
while $S_{y}$, $S_{z}$ decay at late times to a value $\sim\frac{C_{y,z}^{2}}{C^{2}}\sim\frac{\gamma_{y,z}^{2}}{\left(g\mu_{B}B^{ex}\right)^{2}}\ll1$.
The typical time scale for this decay is controlled by the random
variations in the rotation frequency $\omega=\frac{C}{\hbar}=\frac{g\mu_{B}B_{ext}}{\hbar}+\frac{\gamma_{x}B_{x}}{\hbar}+O\left(\gamma^{2}\right)$.
The central spin's purity $\left|S(t)\right|=\sqrt{S_{x}^{2}(t)+S_{y}^{2}(t)+S_{z}^{2}(t)}$
decay time in the limit of high external field is thus given by
\[
T_{2}=\frac{\hbar}{\gamma_{x}},
\]
or in other words the purity decay time is inversely proportional
to the hyperfine tensor component which is parallel to the externally
applied magnetic field.

\bibliographystyle{aipnum4-1}